# Of (Biological) Models and Simulations

Maurice HT Ling[1,2]*

[1]Colossus Technologies LLP, Republic of Singapore
[2]School of BioSciences, The University of Melbourne, Australia

*Corresponding author: Maurice HT Ling, Colossus Technologies LLP, 8 Burns Road, Trivex, Singapore, 369977, Republic of Singapore, Tel: +65-96669233; Email: mauriceling@colossus-tech.com

**Abstract**

Modeling and simulation are recognized as important aspects of the scientific method for more than 70 years but its adoption in biology has been slow. Debates on its representativeness, usefulness, and whether the effort spent on such endeavors is worthwhile, exist to this day. Here, I argue that most of learning is modeling; hence, arriving at a contradiction if models are not useful. Representing biological systems through mathematical models can be difficult but the modeling procedure is a process in itself that follows a semi-formal set of rules. Although seldom reported, failure in modeling is not a rare event but I argue that this is usually a result of erroneous underlying knowledge or mis-application of a model beyond its intended purpose. I argue that in many biological studies, simulation is the only experimental tool. In others, simulation is a means of reducing possible combinations of experimental work; thereby, presenting an economical case for simulation; thus, worthwhile to engage in this endeavor. The representativeness of simulation depends on the validation, verification, assumptions, and limitations of the underlying model. This will be illustrated using the inter-relationship between population, samples, probability theory, and statistics.

## Models and Simulations

All models are wrong, but some are useful.

Box and Draper [1].

All models are right… Most are useless.

Tarpey [2]

The role of models has been recognized as an important aspect to the scientific method [3] for more than 70 years [4]. In recent years, there are increasing interests in metabolic modeling in biology [5,6]. The process of building a model is known as modeling. Generally speaking, models can be classified as either material models or formal models. Rosenblueth & Wiener [7] consider material models to be a physical representation of the object under study whereas formal models are symbolic or logical representations of the object under study. For example, a scale down model of an airplane for testing its aerodynamic properties in a wind tunnel is a material model. The main purpose of a scaled-down material model is the ease of study, which had been clearly demonstrated by a study on fireplace more than 200 years ago [8].

A formal model, on the other hand, is an abstract but unambiguous description of a physical object or phenomenon, usually in logical or mathematical constructs. Commonly used mathematical constructs for building formal models are rate equations, such as differential equations [9], or state transitions, such as petri nets [10]. Other modeling formalisms had been reviewed by Machado [11]. Besides differentiating models by their underlying formalism, models can also be classified by their informational complexity. At the lowest level, the presence or absence of associations between 2 entities (such as cells or molecules) can be represented as a Boolean term. This results in a binary state petri net or Boolean network, which can be used to study cycles and reachability. For example, a metabolic pathway or gene regulation network is essentially a Boolean network [12]. At the next level, the stiochiometries between each state can be added [13], resulting in a steady state model, which can be simulated and used to study the state or concentrations of various molecules at homeostasis using methods such as flux balance analysis [14-16]. At the highest level, the transitions between each state can be represented using a set of differential equations and this result in a kinetic model [17]. The main advantage of kinetic models over steady state model is the ability to simulate each state (or molecules in the context of metabolic modeling) across time, in addition to the ability to model homeostasis [12]. This will allow for identification of bottlenecks in the metabolism [18].

With advancements in computing power, formal models, being constructed in mathematical or logical format, render it possible for computation. Such computation or execution of formal models is commonly known as simulation. Hence, by definition, one can "build a model", then "run or execute a model" or "run or execute a simulation" but in no means, one can "build a simulation". The ability to render execution of a model; thus, an executable or simulatable model; defines the fundamental difference between a material model and a formal model. In addition, there are two kinds of modeling strategy in the current literature [19] forward and reverse modeling. Reverse modeling starts from experimental data and seeks potential causalities suggested by the correlations in the data, captured in the structure of a mathematical model. Forward modeling starts from known, or suspected, causalities, expressed in the form of a model, from which predictions are made about what to expect.

Thus, simulation can be seen as the utilization of a formal model. Very commonly, the purpose of building a formal model is to simulate. As such, in this article, I will use the term "model building" to specifically refer to the process of building a formal model, "simulation" to specifically refer to the process of simulating a formal model, and "modeling" to refer to the both the process of model building and simulation.

The most important purpose of modeling is to gain insights into behaviors that are not obvious from the model itself [20] by reading the mathematical equations. In spite of this, there is an ongoing debate on the role of models and simulations in biology and science at large, even to this day. The major arguments against modeling in biology can be summarized in three questions:

i. Is a model/simulation useful [3]?

ii. Is a model/simulation representative [21]?

iii. Is it worth spending the time and effort engaging in modeling [22]?

For the rest of this article, I will attempt to answer these questions and take the view that modeling is both useful and can be representative; hence, it is worth the time and effort to engage in this endeavor.

## All Models are Wrong but some are Useful

Are models useful? In order to answer this question, we have to ask, what are the uses of models? Thus, if models exhibit its required utility, models are; therefore, useful. To start off, I argue that models are an integral part of learning. Learning from models and learning by models are collectively known as model-facilitated or model-based learning, which is the formation of mental models of knowledge by the learner [23,24]. In a classical monograph titled "What the Best College Teachers Do" [25], Ken Bain pointed out that knowledge from teacher to student is not transferred but built in the student and the process of doing so is likened to assisting the student to build a mental model of the knowledge, followed by reasoning using the built mental model. This is then followed by critically examination of the model and the role of the teacher is to bring the student's mental model to a point of "expectation error", where the mental model of knowledge breaks down, which is analogous to finding the limitations of models. As Bain [25] puts it:

"The best college and university teachers create what we might call a natural critical learning environment in which they embed the skills and information they wish to teach in assignments (questions and tasks) students will find fascinating – authentic tasks that will arouse curiosity, challenge students to rethink their assumptions and examine their mental models of reality."

Here is a personal experience when I was trying to learn about DNA replication. As DNA is double stranded and replication can only occur from a 5' to 3' direction, it means that one strand of the DNA (known as the lagging strand) has to be "copied backwards" from a seemingly 3' to 5' direction, which is a contradiction. Work by Sakabe & Okasaki [26] showed that the lagging strand is actually copied in short fragments in a 5" to 3" direction, which eventually came to be known as Okazaki fragments. The primers of Okazaki fragments are subsequently removed by a combination of enzymes. However, I could not understand why the enzymes did not zip pass and remove the entire downstream Okasaki fragment instead. My instructor realized this contradiction and suggested that the chemical structure of ribose (in the primer) and deoxyribose (in the DNA section of Okasaki fragment) may play a role [27]. Hence, my mental model of DNA replication was correct but incomplete as it did not take into account of the chemical structure between ribose and deoxyribose, which lead to an expectation error (why the entire downstream Okasaki fragment was not removed?). This led me to modify my mental model.

The advancement of science depends on constant challenging and re-examination of perceived notions and assumptions, and to incorporate new information from latest research work into existing models. In the process of doing so, mental models of knowledge are refined and used as basis for extension, making model a powerful tool for learning scientific knowledge and reasoning [28]. Despite the advantages of using models in learning [25,28], there are perceived notions that modeling may not be useful as models may be too specific and difficult to generalize [20,29]. These notions may be indirectly refuted by recent efforts promoting the usefulness of modeling in biology [30,31] and by a large volume of studies in other fields [32]. Underpinning the usefulness of modeling is the balance between specificity and generalization.

I will further argue that model building is perennial in all fields of science on the basis that the simplest model, a single mathematical equation, is a codified model between the inputs and outputs. For example, Newton's Second Law of Motion states that force is proportional to the product of mass and acceleration, is a formal model between force, mass, and acceleration. Similarly, in microbiology, Monod's equation is a century old equation which formalizes the relationship biomass to the concentration of limiting substrate, and is still in use in many studies today [33,34].

Hence, models are meant to be specific - the intent of a model is for one and only one purpose. For instance, a scale model of an airplane for aerodynamic testing in a wind tunnel will not be applicable for propulsion testing. If the same scale model can be used for another purpose not in its original specification or intent, it is a bonus and not a requirement. Conversely, if the same model cannot be used for another purpose other than its original intent, it is expected rather than a criticism. Thus, all models are wrong except when used for its intended purpose. This is also related to the choice of modeling techniques. For example, if the requirement is to identify bottlenecks in redox reactions, then it is necessary to build a kinetic model.

To push this argument further, it can also be said that a model need not be correct in all aspects. In fact, the areas in which the model is wrong may be generalized as the limitations of the model. For example, Mendelian genetics is a model of inheritance which had been used for more than a century as the basis for studying inherited disorders [35]. Despite so, instances of inheritance that fail to follow Mendelian genetics are widely known [36]. Yet, this does not mean that either model is wrong. In fact, both models are correct in a limited way. Thus, all models are wrong except when used for its intended purpose and with limitations. Indeed, even the seemingly ubiquitous Monod's equation [37,38] in microbiology has its own limitations. The same can be said for Newtonian mechanics, which is known to be irrelevant when the speed of the object is near light speed.

In fact, a model need not be highly accurate to be useful. For example, Karr, et al. [39] constructed a whole cell metabolic model of *Mycoplasma genitalium*. Although after modeling all 525 genes and its metabolism, Karr, et al. [39] managed to achieve a correlation of 0.82 (reported as R$^2$ of 0.68) between observed experiments and simulated results. In the strictest sense, Karr, et al. [39]'s model can be deemed to be "having room for improvements" as the correlation is not near perfect (that is, $R^2 > 0.95$) but this did not prevent the model to be used to gain insights into various metabolic processes of *M. genitalium*. This suggests that a model only needs to be sufficiently accurate or correct in some aspects but not all. This work [39] had even been commented in the same issue of the journal as the "dawn of virtual cell biology" [40]. Such large scale models had been attempted in other organisms and had provided insights [41] even for industrial applications [42]. Larger scales of models of various single cell models had been attempted for studying the emergence of microbial ecology [43] and it is not expected for these models to be entirely highly to be useful. Hence, there is no practical reason to be hung up on high accuracy of models, especially when it is generally known that laboratory experiments may have varied degrees of reproducibility due to inaccuracies of instruments (such as, imprecision of micropipetters) and skills of the researcher (such as, not adhering to protocols down to the seconds). In many cases, the purpose of the models is to provide insights, which have to be experimentally validated, but nevertheless drive research directions.

Albert Einstein is reported to have said to the effect that "*no amount of experimentation can ever prove me right; a single experiment can prove me wrong*" [44]. Science should be worried about models without limitations, and by extension, worry about models that are always right. This will only imply that such models have not been carefully examined. Thus, all models have to be wrong in some ways to render its greatest use.

At the end of the day, science had always been in the business of proving models to be wrong and that is the basis of hypothesis testing. Personally, I had threw away more models than I am willing to count. Like a tree today is sitting on a mass grave of past plants and animals, the models that I have today are standing on a graveyard of models. Yet, all these dead or failed models, which are common place and un-publishable, taught me something about my own understanding of the field and more importantly, highlighting my mis-understanding of the field.

### Modeling, the Only Experimental Tool

In the most basic form, a model can be seen as the box or system between the input or stimuli, and the output or response [45]. The main goal of science is to understand and explain how the response(s) is/are a result of the input(s) [46]. Taking Monod's equation as an example, the concentration of limiting substrate is the input to calculate the biomass, and the equation itself explains the logical relationship between the concentration of limiting substrate and the biomass.

This implies that there are 3 uses of models [45]. Firstly, by knowing the input and the model, the output or response can be predicted. Secondly, by knowing the model and the required response, the input can be controlled. Lastly, by knowing the input and response, the system can be understood.

However, these 3 uses are not independent. Very often, a model has to be relatively understood before it can be used for prediction or control. If a model is not substantially understood, it often gives wrong predictions and the model has to be refined and improved upon. Scientific experiments can then be seen as a procedure to generate a set of corresponding input-output pairs on which the underlying phenomenon can be understood; thus, modeled.

Although experimentation in biology is common (as evident by the large volume of experimental publications), laboratory experimentation is both difficult and unethical in many areas of biology [47], such as epidemiology and evolution. Real-world epidemiological experiments will require the willful dissemination of infectious agents into the general public as the input to track the speed and routes of transmission as the output. Clearly, this is not ethical and should never be allowed. Hence, most of epidemiology is the collection of data on existing infections from the general public [48] for the purpose of modeling the mode of infection [49] using methods such as curve fitting techniques [50]. Only then, can outcomes of disease transmission be studied and tested in a virtual context before using it for forecasting. Similarly, studying evolution experimentally is difficult and expensive [51]. Although some experimental evolution studies have been done [52-55], they are restricted to evolution of fast-growing bacterial cells. It is impossible to study human evolution in an experimental setting, both in terms of ethics and in duration.

In these ethically and/or practically impossible situations, modeling appears to be the only viable experimental tool. Computer simulations of virtual organisms (commonly known as "digital organisms" or "artificial life") had been used instead (reviewed in [56]) and may provide some insights into human evolution [57]. Moreover, there are areas that intersect between evolution and epidemiology, such as the evolution of antibiotics resistance. There have been contradictory studies showing that antibiotics resistance can possibly revert once the specific antibiotic is dis-used [58]. Although antibiotics resistance had been studied experimentally [59], it is only performed in controlled laboratory settings as it is unacceptable to willfully induce one or more antibiotics resistance in the human or animal population at large. Hence, modeling of virtual organisms has been used as an experimental tool in attempt to break the above contradiction and find that contradictory results are caused by statistical variations in the de-evolution of antibiotics resistance as the 95% confidence interval can vary from reversion to non-reversion of antibiotics resistance [58,60].

Hence, the question of whether models are useful appears to be rather muted in situations whereby models are the only feasible instruments of study.

### Modeling, the Economic Tool

It is without a doubt that biology is complex and there can be a lack of thorough understanding in underlying biology or the lack of representative data but Gunawardena [19] argues that these limitations do not warrant the call for not modeling. I will further argue that such limitations, especially the lack of representative data, may be what render modeling useful – to build an initial model using available data, and using the model to uncover what data is needed to improve and refine the model. Using the analogy of learning a new language - nobody attempt to learn a new language by memorizing all the grammatical rules and dictionary of words prior to constructing a simple greeting. Recently, Chowdhury, et al. [61] had demonstrated the use of models as a means to consolidate on-going research results.

Modern science is not just complex but expensive to study experimentally. Multi-billion dollar scientific equipment, such as the Hubble telescope and Large Hadron Collider, had been built. Biology is of no exception as it can often take millions of dollars to build a modern biology laboratory. This has been suggested to be widening a disparity between wealth of different research groups or even countries [47]. Modeling may present itself as an economic leveling tool in this aspect, as well as a tool towards efficient use of research dollars.

There have been a number of recent metabolic engineering studies demonstrating the use of modeling to reduce the number of experiments needed [62] to optimize specific metabolite production [63,64]. Thus, resulting in more efficient use of research dollars as a result of higher research and development throughput and is likely to result in shorter time between research and industrial production.

Machado, et al. [63] modeled a 4-enzymatic step for curcumin production from phenylalanine and/or tyrosine, and carried out *in silico* modeling experiments to isolate mutants with potential for high curcumin production. Of the possible enzyme over-expressions, Machado, et al. [63] suggested that over-expression of acetyl-coA carboxylase, which catalyzes the acetyl-coA to malonyl-coA, might have the biggest impact in terms of increased curcumin production. This result is un-expected as acetyl-coA carboxylase is not the first enzyme in the sequence that utilizes phenylalanine or tyrosine, nor the enzyme directly responsible to curcumin synthesis; hence, suggesting a non-intuitive result.

Weaver, et al. [64] modeled a 6-enzyme pathway from mevalonate to amorphadiene with the main of modifying enzyme kinetics, such as turnover number and Michaelis-Menten constant of the enzymes as well as concentrations of enzymes, in order to maximize amorphadiene production. Hence, Weaver et al. [64] went one step further compared to Machado, et al. [63] as Weaver, et al. [64] is interested to which property of the enzyme (turnover number and/or Michaelis-Menten constant) is more important. Through sensitivity analysis, Weaver, et al. [64] found that the turnover number and expression level of amorphadiene synthase, which catalyzes the last step to product amorphadiene, are important in the final production rate of amorphadiene. However, the Michaelis-Menten constant of amorphadiene synthase has negligible impact on amorphadiene production rate. Weaver, et al. [64] validated this modeling findings experimentally.

These recent studies [63,64] demonstrate the use of modeling to reduce the number of optimization trials to run in order to get a potentially economically bacterial strain for industrial fermentation. They also [63, 64] further underline a usefulness of modeling as a window to access the system on hand [65]. This view is supported by Lung [66], whom suggests that models are useful as "*models can provide insight into the behavior of complex systems and sometimes yield results which are different from the intuitive predictions.*" Taken together, modeling may be a potential tool for improved stewardship of limited research funding.

## Simulacrum

Recently, there are also reported works using modeling as a tool for clinical settings [67,68], suggesting that efforts spent to learn about modeling is worthwhile [67]. This brings us to the last issue – what makes a model representative and by extension, how interpretable are the simulation results? These appear to be new philosophical questions arising from modeling but Frigg & Reiss [69] argue that these seemingly new philosophical questions have analogies in experimentation and thought experiments; hence, they are not new questions but a rehashing of existing questions.

I am inclined towards Frigg & Reiss [69]. In the field of biology, the term "model" has been used in many different contexts to refer to an experimentation platform (such as model organism) or an abstraction (such as fluid-mosaic model of cell membrane) and the question of representativeness had been addressed according; for example, "is zebrafish a good model for developmental biology [70]?" Unless we want to use humans as model organism for human development, we are not left with much choice but to use appropriate experimental models; which balance practicality, scientific value, and ethics; as well as acknowledging the limitations of such models. However, this may also be a fatalistic point of view.

Another way to view this is using the relationship between population and sample as an analogy. Here, population refers to the object of interest but is practically impossible to study in its entirety. On the other hand, a sample is a reduced population or a representative population, which is analogous to a model and is feasible for study. Then, what makes a sample a representative population? The answer will be "probability theory" - every reasonable feature in the population has to be proportionally represented in the sample. It is important to note that it may not be possible to represent every single feature from the population onto the sample. Very often, features of low probabilities get left out and this has to do with sample size. For example, if the side-effect of a drug in question happens 1 in 100 thousand, it is unlikely to be represented in a sample of 10 thousand patients administered the drug. This forms the inherent limitation of the sample, and by extension, the model. After analyzing the sample, the analysis results have to be interpreted in the context of the population. This is the work of "statistical theory". Within this interpretation, the sampling limitations must be reflected.

Here, I argue that the inter-relationship between population, samples, probability theory, and statistics, forms the basis of simulacrum. At the fundamental level, the simile to probability theory and statistics, when applied to modeling, are verification and validation, respectively.

In layman's terms, verification and validation can be asked in the following questions:

i. For verification, "are we building the model right?" That is, is the model free from implementation errors and/or logical errors [71]?

ii. For validation, "are we building the right model?" That is, is the model addressing the needs and requirements [71]?

In essence, verification is concerned with internal or within-model "correctness" while validation is concerned with external "correctness" [72].

Using the work of Massoud, et al. [73], verification can be performed at 2 stages: model building stage, and internal verification of the model. In building stage, the real-world system (analogous to population) is formulated into a model (analogous to sample) via inductive generalization, also known as semantic link. This calls for the identification of a set of minimum parameters used to describe the real world system and to form associations based on observations in the real world. This results in a specific model, which has to be verified internally. In this stage (model verification), the main purpose is to ensure that the semantic associations between each parameter is logically meaningful based on the current knowledge in the field. By Occam's Razor, simple associations are preferred over complex associations, and a set of small number of associations is preferred over a larger set.

Once all associations are verified, testing of the model is performed for accuracy by matching the model's behaviors with the observations in which the model is built from. This step is mandatory as it determines the degree in which the model is an accurate representation of the system under study [3,20]. It is common in mathematical models that stochastic variations arise from imprecisions in floating-point calculations, which are commonly known as rounding-off errors. Hence, it may be important to consider whether such errors are within acceptable range and whether a change of scales may reduce such errors. For example, 1% error in molarity represents 10 milli-moles whereas 1% error in milli-molar represents 10 micro-moles. In addition, the importance of each parameter can be further estimated using sensitivity analysis [74] and it may be possible to eliminate parameters of low importance for the purpose of simplifying the model. Several iterations of model refinements and testing are usually carried out to produce an 'adequate' model that is verified with acceptable accuracy. The model can then be said to be representative of the system under study.

Only then, can the model be validated. Validation can be done in primarily 2 ways, as described by Hicks, et al. [75]; namely, (1) comparing simulation results with additional observations, and (2) comparing simulation results with that of other validated models.

When comparing simulation results with additional observations, it is important to note that these observations cannot be used for model building; implying that a set of new observations must be used for validity testing. A commonly used technique is known as cross validation [76], also known as rotational estimation. In cross validation, the entire set of observations is randomly divided into 2 or more equal sub-sets where 1 sub-set is reserved for validation testing and the remaining sub-sets are used for model building. The error or accuracy of the model is then reported as average (mean of means) and standard error.

In event where other validated models are available, it is also common to compare results between 2 or more models. However, there are 2 common issues with this approach. The parameters used in other models may be different from the set of parameters used in the current model, which may be a result of different assumptions and considerations during model building. The purpose of the models may also be different. This is likely to result in differences in simulation outputs as comparison should be made only when 2 models uses the same set of parameters. The differences in the simulation output will be confounded with the difference in parameters used, giving another obstacle in interpretation.

Besides the above 2 methods, Sargent [71] encourages the elucidation of operational validity to determine whether the model's output has the accuracy required for its intended purpose. A major aspect of operational validity is accuracy, which has been dealt with in the above verification and validation. The glaring aspect that had not been addressed is the limitations of the model, which can be defined as the boundaries of the applicability of the model. In this case, extreme parameter conditions have to be tested to estimate the conditions in which the model fails; thus, setting operational boundaries for the model. Within these operational boundaries, the model is able to give insights into the system under study (analogous to statistics providing insights into the population).

## Conclusion

Modeling has been an important aspect of science [71,77] but there is an ongoing debate on the role of modeling in biology and science at large, even to this day. In this article, I argue that modeling can be useful tool [72,78] to aid in many areas of biological studies. The use of modeling extends beyond being a research tool but may be an economical tool to reduce the number of experimental trials needed, which may lead to a better use of limited research funding. Thus, it is worth spending effort to learn and incorporate modeling into a biologist toolkit.